\documentstyle[twoside,fleqn,espcrc2]{article}
\newcommand{\AmS}{{\protect\the\textfont2
  A\kern-.1667em\lower.5ex\hbox{M}\kern-.125emS}}

\title{Introduction to D-branes
}

\author{L\'arus Thorlacius\address{Department of Physics, 
        Princeton University, \\ 
        Princeton, NJ, 08544, USA}%
        \thanks{Lectures presented at the 33rd Karpacz Winter School 
	of Theoretical Physics, Karpacz, Poland, 13-22 February 1997 }
       } 

\def\be{\begin{equation}}
\def\ee{\end{equation}}

\def\ba{\begin{array}}
\def\ea{\end{array}}

\def\NS{NS-NS}
\def\RR{R-R}

\input epsf.tex
       
\begin{document}
\bibliographystyle{unsrt}

\begin{abstract}

D-branes are string theory solitons defined by boundary 
conditions on open strings that end on them.  They play
an important role in the non-perturbative dynamics of
string theory and have found a wide range of applications.
These lectures give a basic introduction to 
D-branes and their dynamics.
 
\end{abstract}

\maketitle

\section{Introduction}

Dirichlet-branes, or D-branes for short,
have moved to center stage in string theory.  In a 
relatively short time they have gone from being esoteric, and
largely ignored, extended objects that existed in certain string
theories to playing a key role in the present understanding of
non-perturbative dynamics and duality in string theory.  One
reason that the importance of D-branes was overlooked is 
that they are associated with open strings and, at the 
perturbative level, open string theory appears much less
promising than closed string theories, such as the heterotic
string theory.  With the development of string duality, attention
shifted towards non-perturbative issues including solitons, and
it was realized that D-branes had precisely the required 
attributes to be partners with fundamental string states under
duality transformations.  This realization, along with the fact
that D-branes are considerably simpler to describe than other
string solitons and often allow explicit calculations in places 
where only speculations went before, led to an explosion of 
activity involving D-branes that is still going strong.  

These lecture notes are intended as a brief introduction to 
D-brane physics and no attempt will be made to review all
their applications.  We will focus on D-branes in the context
of the type~II oriented superstring theory. 
We begin with a discussion of p-brane soliton solutions of 
type~II supergravity theory, the low-energy effective field 
theory of type~II superstrings.  We then introduce 
Dirichlet-branes.  They are string solitons defined in terms of
open strings that end on them.  The worldsheet field theory of
these open strings is a simple two-dimensional boundary 
field theory and has exact conformal symmetry, as required by
the classical equations of motion of string theory.  Using
worldsheet techniques, we show that D-branes carry the
appropriate conserved charges in spacetime to be the string
theory realization of some of the p-branes of the effective
field theory.  In the remainder of the lecture notes, we explore
some basic properties of D-branes and their dynamics.

\section{Solitons in Type~II Supergravity}

Our starting point for discussing string solitons is the low-energy
effective field theory  that describes strings on length scales that
are large compared to the fundamental string length 
$l_s=\sqrt{\alpha'}$.  In the case of type~IIA and IIB string theories
the effective field theory is $D=10$ supergravity of type IIA and
IIB, respectively.  The label II refers to the fact that these theories
have two independent supersymmetries and the A and B refer to
whether the two ten-dimensional supercharges have the opposite
(A) or the same (B) chirality.  In the string theory this difference 
corresponds to a freedom of choice between two consistent
GSO projections that remove the tachyon from the string 
spectrum.  These theories contain a number of massless bosonic
fields and their fermionic superpartners.  The bosonic fields are
sufficient for our purposes here.  The type IIA and IIB supergravities
have in common a symmetric metric tensor $g_{\mu\nu}$, an
anti-symmetric rank two tensor potential $b_{\mu\nu}$, and a
scalar dilaton $\phi$.  In string theory these fields are associated
with worldsheet fermions with anti-periodic (Neveu-Schwarz)
boundary conditions in both the left- and right-moving sectors
of the closed string, and hence they are commonly referred to as
\NS\ fields.  The remaining bosonic fields in
the supergravity theory are associated with periodic (Ramond)
worldsheet fermions.  In the type IIA theory the \RR\ fields are a
vector potential $a^{(1)}_\mu$, and an anti-symmetric rank three
potential $a^{(3)}_{\mu\nu\lambda}$, while in the IIB theory they 
are a scalar $a^{(0)}$, and anti-symmetric potentials of rank two
and four, $a^{(2)}_{\mu\nu}$ and $a^{(4)}_{\mu\nu\lambda\sigma}$.

The bosonic terms in the effective action for type IIA supergravity
are
\begin{eqnarray}
S_{IIA} &=&  {1\over 2\kappa^2} \int d^{10}x \, \sqrt{-g}
\Bigl[e^{-2\phi}
\bigl(R+4(\nabla\phi)^2 \nonumber \\
&  & -{1\over 2\cdot 3!}H^2 \bigr)  
-\bigl({1\over 4}{h^{(2)}}^2 
+ {1\over 2{\cdot}4!}{h'^{(4)}}^2 \bigr) \Bigr] \nonumber \\
&  & -{1\over 4\kappa^2} \int h^{(4)}\wedge h^{(4)}\wedge b  \>,
\end{eqnarray}
where $H=d\wedge b$ is the \NS\ antisymmetric tensor field
strength, $h^{(n)}= d \wedge a^{(n-1)}$ are \RR\ field strengths,  
and $h'^{(4)} = h^{(4)}+a^{(1)}\wedge H$.  
The \NS\ sector terms come with a prefactor of 
$e^{-2\phi} \sim {g_s}^{-2}$ reflecting that
they arise at lowest order in string perturbation theory.
The \RR\ sector terms, on the other hand, do not have
such a prefactor.  This can be traced to
the structure of the \RR\ vertex operators 
(see for example \cite{pol2})
and ultimately to the fact that fundamental string
states do not carry \RR\ gauge charges.

The bosonic part of the type IIB supergravity action is 
\begin{eqnarray}
S_{IIB} &=&  {1\over 2\kappa^2} \int d^{10}x \, \sqrt{-g}
\Bigl[e^{-2\phi}
\bigl(R+4(\nabla\phi)^2 \nonumber \\
&  & -{1\over 2\cdot 3!}H^2 \bigr)  
-{1\over 2}(\nabla a^{(0)})^2 \nonumber \\
&  & -{1\over 2\cdot 3!}(a^{(0)}H+h^{(3)})^2 \Bigr] \>, 
\end{eqnarray}
plus some terms involving the self-dual $h^{(5)}$, which
cannot be written in covariant form.  The equations of motion
involving $h^{(5)}$ are nonetheless covariant and can be deduced
from the $D=10$ supersymmetry algebra \cite{schwarz}.

These theories have classical solutions 
which describe extended objects called p-branes.  
A p-brane extends
in p spatial directions and traces out a p+1 dimensional 
worldvolume in spacetime.  The p-branes are topological 
solitons carrying conserved charges that act as sources 
for the various anti-symmetric gauge fields.  

The \NS\ sector is the same in the IIA and IIB theories,
and includes a three-form field strength $H$.  
Recall that a point charge in electrodynamics is a source for
a two-form field strength.  This generalizes to higher-dimensional
objects; a p-brane carries the charge of a (p+2)-form field
strength.  The `electric'
charge that couples to $H$ is thus carried by a 1-brane.  The 
corresponding `magnetic' field strength is a seven-form,
\be
\tilde H = e^{-2\phi} *H \>,
\ee
where $*H$ is the Hodge dual of $H$.  
The magnetic charge that couples to $\tilde H$ is carried
by a 5-brane.  

There is a two-parameter family of black string solutions
labeled by the electric $H$ charge and the mass per unit 
length \cite{horstr}.  The term `black' refers to the fact 
that for generic values of the mass the string is surrounded
by an event horizon.  At the extremal value, where the mass
equals the charge in appropriate units, the event horizon 
shrinks away and we are left with an extended fundamental
string \cite{dabhar}.  The supersymmetry algebra dictates
that the mass per unit length of a black string must always
be larger than or equal to the charge per unit length, and 
the fundamental string saturates the inequality.

Similarly there is a two-parameter family of black 5-brane
solutions \cite{horstr}, labeled by the mass and magnetic
charge per unit five-volume.  The supersymmetry algebra
again implies a lower bound for the mass, $M \geq Q$,
which is saturated
by the extremal \NS\ 5-brane described in \cite{dulu,chs}.
The mass bound is a generic feature of p-brane solutions.
The configurations that have equal mass and charge per p-volume
are referred to as BPS states (short for 
Bogomolnyi-Prasad-Sommerfield).  
They are annihilated by some fraction of 
the supercharges of the theory and as a result they sit in
short multiplets of the supersymmetry algebra \cite{oliwit}.
If the theory has enough extended supersymmetry the mass and
number of short multiplets in the spectrum is not changed by
quantum corrections, even at strong coupling.
BPS states have therefore 
played a key role in the study of strong/weak
coupling duality in gauge theory and string theory in recent
years.

The p-branes of most interest to us in these lectures are 
associated with the \RR\ sector fields.  The IIA theory has
0-branes that couple to the two-form field strength 
$h^{(2)}$ and 2-branes that couple to $h^{(4)}$.
Then there are 4- and 6-branes that couple to the 
magnetic fields $*h^{(4)}$ and $*h^{(2)}$, respectively.
In addition one can consider 8-branes.  The corresponding
ten-form field strength is not dynamical, but a constant
field corresponds to a physical energy density \cite{rom,pol}.

In type IIB theory there are p-branes with  p=-1,1,3,5,7. 
The p=-1 case describes instantons, which are events in Euclidean spacetime,
and couple to the field strength of the \RR\ scalar $a^{(0)}$.  
The \RR\ 1- and 5-branes couple to the electric and magnetic field 
strengths $h^{(3)}$ and $\tilde h^{(7)} = * h^{(3)}$, respectively.  
The 3-brane acts as a source for the self-dual field $h^{(5)}$ and the 
7-brane couples to the magnetic field $\tilde h^{(9)} = * (da^{(0)})$.  
There are no 9-branes in type II supergravity.  The corresponding
field strength would have to be of rank eleven and must therefore 
vanish in ten dimensions.  In type I string theory, however, 
9-branes play an important role in giving rise to the $SO(32)$
Chan-Paton gauge degrees of freedom \cite{pol2}.

The black p-branes of type IIA and IIB supergravity are
approximate classical solutions of string theory valid for sufficiently
large mass per volume.  Exact classical solutions of string theory
are described by two-dimensional superconformal field theories.
The \NS\ sector 1- and 5-branes can be systematically
studied using two dimensional sigma models with the 
appropriate field content and couplings \cite{dabhar,cs},
but this approach is not well suited for \RR\ 
sector backgrounds.
Fortunately there is another way to proceed.
It involves a somewhat indirect construction in terms of auxiliary
open strings but, as we shall see below, the resulting two-dimensional
conformal field theory is very simple and the extended objects
so defined indeed carry \RR\ charge and have the right
mass density required by string duality.

\section{Dirichlet-Branes}
In order to describe \RR\ solitons in type II string theory we introduce
open strings into the theory.  We then have to prescribe boundary
conditions for the worldsheet fields at the worldsheet boundary.
The bosonic worldsheet fields $X^{\mu}$, for $\mu=0,\ldots, 9$, are
interpreted as the embedding coordinates of the string in spacetime.
Let them satisfy
\begin{eqnarray}
(N)& n^a \partial_a X^m = 0  & m=0, \ldots, p, \nonumber \\
(D)&  X^i = {\rm const} & i=p+1, \ldots,9, 
\label{dbc}
\end{eqnarray}
where $n^a$ is normal to the boundary.  The N and D refer to 
Neumann and Dirichlet boundary conditions. Conventional open
strings in type I string theory have Neumann boundary conditions for all
the $X^{\mu}$ fields.  With the above prescription the open strings are
constrained to end on a p+1 dimensional hypersurface in spacetime
but the ends are free to move on this surface, see Figure 1.  
\vskip 0.5cm
\vbox{
{\centerline{\epsfxsize=1.5in \epsfbox{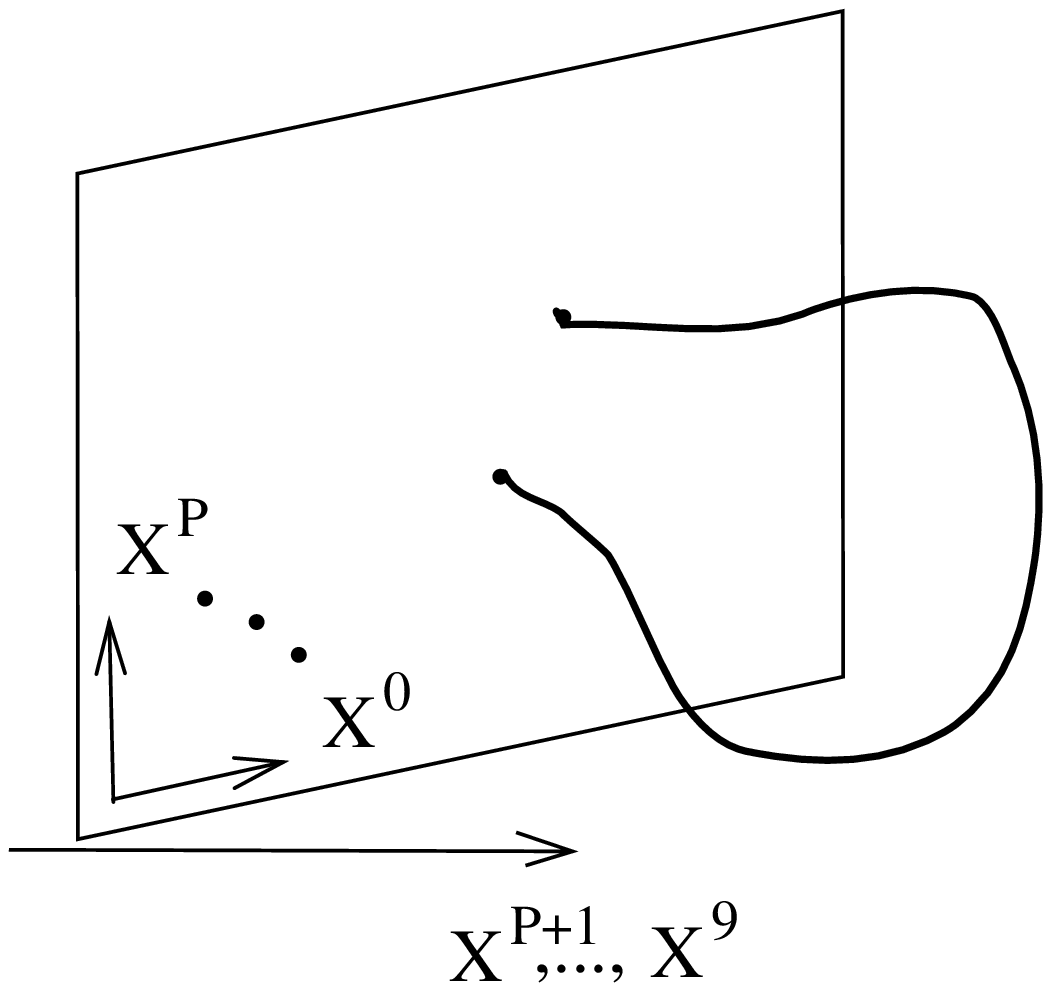}}}
{\centerline{ FIGURE 1:
Dirichlet p-brane with open }}
{\centerline {string attached.}}
}
\vskip .5cm
This defines the p+1 dimensional
worldvolume of an extended object, called a Dirichlet-brane.  At this stage
the D-brane appears to be a rigid sheet infinitely extended in
p spatial directions.  Later on we shall see that a D-brane is in fact dynamical
object with certain massless open string modes describing its
undulations.  A D-brane can be given a finite volume in a spacetime with
compact dimensions by wrapping around non-trivial cycles.  
Finite volume D-branes
that can contract to a point are presumably unstable and are not described
by simple classical solutions.  

We also need to prescribe boundary conditions for the worldsheet fermions. 
The surface terms in the fermion equations of motion cancel if we
let the left-and right-moving fermions, $\psi^\mu$ and $\tilde \psi^\mu$,
reflect into each other up to a sign, at the worldsheet boundary.
Since the fermion fields are only defined up to an overall phase we
can set $\psi^\mu = \tilde \psi^\mu$ at one boundary, for example at
$\sigma = 0$.  The sign choice at the other end 
must be correlated with the N and D boundary
conditions on the bosonic fields (\ref{dbc}) in order to preserve worldsheet
supersymmetry, 
\begin{eqnarray}
(N)& \psi^m =\pm \tilde \psi^m  & m=0, \ldots, p, \nonumber \\
(D)& \psi^i =  \mp \tilde \psi^i & \>\> i=p+1, \ldots,9.
\label{fbc}
\end{eqnarray}

In flat spacetime the type II worldsheet field theory 
contains only free bosons and
fermions and is manifestly a conformal field theory.
The linear boundary conditions (\ref{dbc}) and 
(\ref{fbc}) preserve the conformal
symmetry and thus define exact classical solutions of string theory.
The construction is indirect but it is very much simpler than the 
non-linear sigma model description of \NS\ solitons \cite{chs2}.

Dirichlet-branes are BPS states. This can be seen as a simple consequence
of the boundary conditions on the worldsheet fields.  The two
supercharges of the type II theory are defined in terms of left- and
right-moving worldsheet currents which get reflected into
each other at the boundary.  Only the linear combination 
$Q_{\alpha} + c_p \tilde Q_{\alpha}$ of the two supercharges
is conserved, were $c_p$ is a phase factor that comes from a parity
transformation in the transverse directions $X^i$ with $i=p+1, \ldots, 9$.
In other words, a D-brane is invariant under half the supersymmetry of
the original type II theory, which is precisely the BPS condition.

\section{D-Brane Tension and Charge}

The tension in a p-brane is by definition the energy per unit 
p-dimensional volume.  In supergravity theory the tension
can be determined by an ADM procedure, considering the
asymptotic falloff of the gravitational field in the directions transverse to
the p-brane.  In string theory one could compute the one-point
amplitude of a graviton vertex operator inserted into
a worldsheet of disk topology with the
D-brane boundary conditions (\ref{dbc}) and (\ref{fbc})
applied.  This amplitude directly probes the gravitational field of the 
D-brane and allows the tension to be read off \cite{dlp}.
The disk is an open string tree diagram and the amplitude therefore
depends on the dilaton as $e^{-\phi}$.  It follows that the D-brane
tension is proportional to ${g_s}^{-1}$.
Similarly, a one-point disk amplitude involving a vertex operator for the
appropriate \RR\ sector gauge field yields the magnitude of
any \RR\ charge carried by the D-brane.

These calculations are straightforward but require careful normalization,
without which one cannot read off the correct mass and charge.  It
turns out to be easier to instead work with a one-loop string amplitude
which involves the exchange of virtual closed strings
between D-branes.  Following Polchinski we consider a pair
of parallel Dirichlet p-branes, separated by a distance $Y$ in a
transverse direction, and compute the annulus graph in Figure 2.
\vskip 0.5cm
\vbox{
{\centerline{\epsfxsize=1.5in \epsfbox{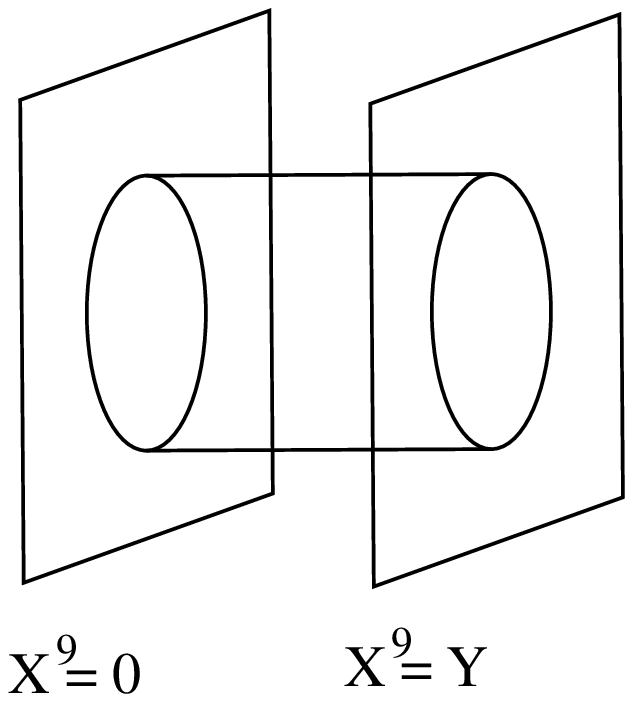}}}
{\centerline{ FIGURE 2:
 Annulus amplitude for open strings}}
{\centerline {stretched between different branes.}}
}
\vskip .5cm
Worldsheet duality allows us to either view this diagram as
an open string vacuum loop with the string endpoints living on the two
separate D-branes or as a tree level closed string exchange of closed
strings between the D-branes, see Figure~3.  
The diagram has a single real valued
modulus which, in the closed string channel, can be taken to be
the worldsheet time $\tau$ between the ends of the cylinder and runs from
0 to $\infty$. In the limit of large $\tau$ the amplitude is dominated
by the exchange of massless closed sting states,
the graviton and dilaton from the \NS\ sector and \RR\ gauge particles.
By identifying the separate contributions from these states
one can read off the tension and \RR\ charge of the D-branes.
\vskip 0.5cm
\vbox{
{\centerline{\epsfxsize=3.0in \epsfbox{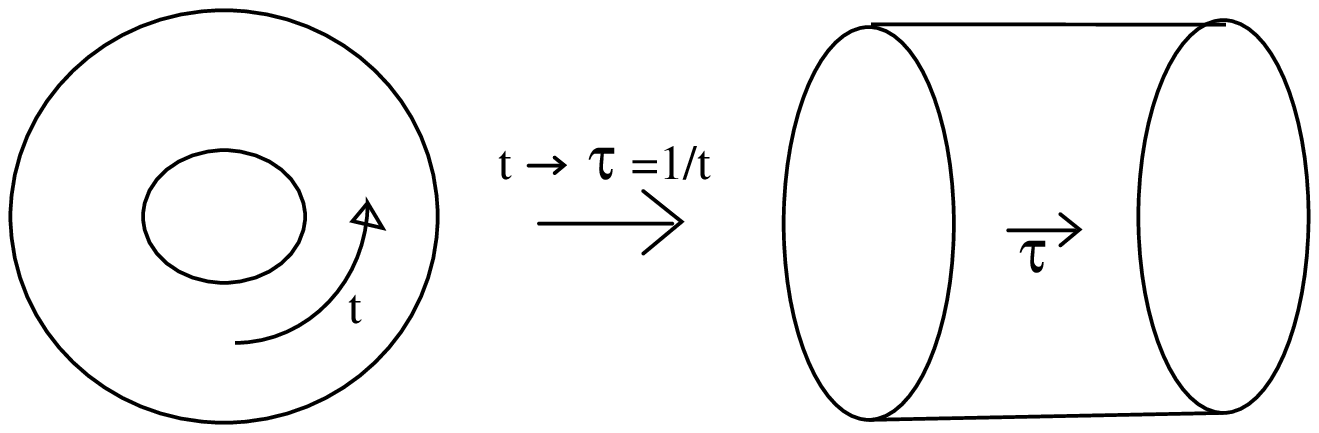}}}
{\centerline{ FIGURE 3:
Open and closed string channels }}
{\centerline {for annulus amplitude.}}
}
\vskip .5cm

To evaluate the diagram we first view it as an open string
loop and write the vacuum amplitude
\begin{eqnarray}
A &=& V_{p+1} 
\int \frac{d^{p+1}k}{(2 \pi)^{p+1}}
 \int_{0}^{\infty} \frac{dt}{2t} \nonumber \\
 & &
\times \sum_{\{n\}} e^{-2 \pi t (k^2 + M_{\{n\}}^2)}.
\end{eqnarray}
$V_{p+1}$ denotes the p-brane worldvolume, the integral is over the zero mode
momentum in the p+1 directions along the worldvolume, and the sum is over the
spectrum of physical modes of an open superstring with ends on separate D-branes.
The mode sum and the momentum integral can be carried out 
explicitly, giving
\begin{eqnarray}
\label{vacamp}
A &=& V_{p+1} \int \frac{dt}{2t} (8 \pi^2 t)^{-\frac {(p+1)}{2}}
e^{- Y^2 t/{2 \pi}} \nonumber \\
& & \times F_0 \> [F_1 + F_2 + F_3].
\end{eqnarray}
The three terms in the brackets are
\begin{eqnarray}
\label{modf}
F_0 (e^{-\pi t}) &=& e^{2 \pi t /3} \prod_{n=1}^{\infty} 
(1-e^{-2n \pi t})^{-8}, \nonumber \\
F_1 (e^{-\pi t}) &=& -16 e^{-2 \pi t /3} \prod_{n=1}^{\infty} 
(1+e^{-2n \pi t})^{8}, \nonumber \\
F_2 (e^{-\pi t}) &=& e^{\pi t /3} \prod_{n=1}^{\infty} 
(1+e^{-(2n-1) \pi t})^{8}, \nonumber \\
F_3(e^{-\pi t}) &=& - e^{ \pi t /3} \prod_{n=1}^{\infty} 
(1- e^{-(2n-1) \pi t})^{8}, 
\end{eqnarray}
and they in fact cancel by Jacobi's famous abstruse identity.  
This is not surprising given
that we are calculating a vacuum amplitude
in a supersymmetric theory. The vanishing of the
amplitude can also be given an
interpretation in terms of closed string exchange.  The exchange of a 
graviton or dilaton gives rise to an attractive force between
the D-branes, while the exchange of a \RR\ gauge boson gives rise to
the repulsive force between two D-branes with the same sign \RR\
charge.  When all the contributions are added
the net force turns out to be zero.
This force balance, which is a familiar feature of BPS objects, only holds
when both D-branes are at rest in some reference frame.
Later on we will discuss the interaction between D-branes when 
one is moving with respect to the other.

Even if the overall amplitude vanishes we can still extract useful information
from the separate terms. The amplitude (\ref{vacamp})
was expressed as an integral over t, the modulus of the
annulus in the open string channel.  The $t \rightarrow 0$ end of the
integration range is an ultraviolet limit from the open string point
of view, but in the closed string channel the modulus
is $\tau = 1/t$ and the corresponding limit is in the infrared.  This is
precisely where the cylinder amplitude is dominated by massless
exchange and we expect to find information
about the tension and \RR\ charge of the D-brane.  In order
to extract the $t \rightarrow 0$ behavior of the
amplitude (\ref{vacamp}) we make use of the
properties of the functions $F_i$ defined in (\ref{modf})
under the modular transformation $t \rightarrow 1/t$
which relates the open and closed string channels,
\begin{eqnarray}
\label{modx}
F_0 (e^{-\pi t}) &=& t^{-1/2} F_0 (e^{-\pi /t}),  \nonumber \\
F_1 (e^{-\pi t}) &=& F_3 (e^{-\pi /t}),  \nonumber \\
F_2 (e^{-\pi t}) &=& F_2 (e^{-\pi /t}).
\end{eqnarray}
Inserting this into (\ref{vacamp}) and expanding out
the infinite products to leading orders one finds
\begin{eqnarray}
A &\approx & (1-1) V_{p+1} \int 
\frac{dt}{t} (8 \pi^2 t)^{-\frac {(p+1)}{2}} \nonumber \\
& & \times 16 t^4 e^ {- Y^2 t/{2 \pi}}.
\end{eqnarray}
Carrying out the t integral one arrives at \cite{pol2,pol}
\be
A \approx (1-1) V_{p+1} 2 \pi (4 \pi^2)^{3-p} G_{q-p}(Y^2),
\label{final}
\ee
where $G_d(Y^2)$ is the radial Green function of a massless
scalar in d spatial dimensions.

We now want to compare this result with a corresponding field
theory calculation.  Consider a theory of a rank $p+2$ field
strength coupled to a p-brane source,
\begin{eqnarray}
S &=& \frac{1}{2(p+2)!} \int d^{10} x {(h^{(p+2)})}^2 \nonumber \\
& & + i \mu_p \int_{V_{p+1}} a^{(p+1)}.
\end{eqnarray}
 The leading order exchange amplitude is easily computed and
has the same value as the \RR\ term in (\ref{final})
provided we make the identification \cite{pol}
\be
{\mu_p}^2 = 2 \pi (4 \pi ^2)^{3-p} .
\label{charge}
\ee
The analogous calculation for graviton and dilaton
exchange in the effective field theory gives, upon
comparing with (\ref{final}), the following D-brane
tension \cite{pol}
\be
\tau_p = \frac {\sqrt{\pi} (2 \pi)^{3-p}}{g_s}.
\ee

Dirichlet-branes are thus seen to be BPS objects
carrying \RR\ gauge charge and have tension
inversely proportional
to the string coupling.  Their definition in terms
of boundary conditions on auxiliary open
strings leads to a two-dimensional worldsheet theory
with exact conformal symmetry as required for classical 
solutions in string theory.  This combination of properties
makes a very convincing case for the identification of
D-branes as the \RR\ sector p-branes of string
theory.  

There is an important check on
the above considerations.  The $p+2$ form field
strength that couples to a p-brane is Hodge dual
to a 8-p form field strength that couples to
a (6-p)-brane.  This is analogous to the duality
between electric and magnetic point charges
in four-dimensional spacetime and
accordingly the brane charge densities must
satisfy a quantization condition analogous to the
Dirac condition \cite{teinep},
\be
\mu_p \mu_{6-p} = 2 \pi n,
\label{dirac}
\ee
where n is some integer.  The \RR\ charges obtained
in (\ref{charge}) satisfy this relation with the minimum
value of $n = \pm 1$.

We conclude this section with some remarks on the relation
between D-branes and the corresponding \RR\ charged solitons
of supergravity theory.  They have the same tension and 
charge density but apart from that they appear quite different.
The D-branes are defined in terms of weakly coupled open strings
that propagate in flat spacetime, whereas a supergravity p-brane
solution describes a spacetime with non-trivial metric and
curvature.  It depends on the value of the string coupling 
which description is more appropriate.  As $g_s \rightarrow 0$,
the gravitational field of the p-brane vanishes and the flat 
space D-brane picture is valid.  I on the other hand, the 
coupling is not small, we are not justified in ignoring the 
gravitational response to the p-brane source and the supergravity
solutions is a more accurate description of the geometry.  
Supersymmetry and the BPS property of extremal p-branes are
crucial ingredients in any argument that takes a result obtained
at weak coupling, such as the number of multi D-brane states 
with some macroscopic charges, and extrapolates it to the strong
coupling regime in order to draw conclusions about a black hole
of the same mass and charge.

\section{Probing D-branes With Strings}
The conformal field theory description of D-branes
is rather simple and allows explicit calculation of
various processes involving D-branes.  For
instance one can compute amplitudes for
closed string scattering off a D-brane
\cite{kletho,ghkm,garmye}.  This provides
information about the structure of D-branes
beyond the classical picture of infinitely thin
hypersurfaces.  The leading contribution to the
p-brane form factor can be read off
from the two-point function of closed-string vertex
operators on a disk with the appropriate mixed
Dirichlet and Neumann boundary conditions.
This corresponds to a string colliding with a D-brane
creating an excitation that propagates along the worldvolume
until the D-brane returns to its ground state by emitting
another string.  For simplicity, we only consider the
scattering of \NS\ sector string states.  The generalization
to \RR\ sector states is straightforward \cite{ghkm}.  The amplitude
takes the form
\be
A = {\langle V_1(z_1, \bar z_1) V_2(z_2, \bar z_2) \rangle}_{Disk},
\ee
where the vertex operator of a \NS\ sector massless state
(graviton or anti-symmetric tensor) with polarization
$\varepsilon_{IJ}$ is
\begin{eqnarray}
V(z, \bar z) & = & \varepsilon_{IJ} 
(\partial X^I + ik \psi \psi^I) \nonumber \\
& & \times (\bar \partial X^J + ik \tilde \psi {\tilde \psi}^J)
e^{ik X(z, \bar z)}.
\end{eqnarray}
For evaluating the amplitude it is convenient to map
the disk to the upper half of the complex plane
and perform Wick contractions using the
following propagators for the worldsheet fields,
\begin{eqnarray}
\langle X^{\mu}(z, \bar z) X^{\nu}(w, \bar w) \rangle & = &
-2 \eta^{\mu \nu} log \mid z-w \mid \nonumber \\
& &- 2 a^{\mu \nu} log \mid 1-z \bar w \mid,
\nonumber \\
\langle \psi^{\mu}(z) \psi^{\nu}(w) \rangle & = &
\frac{ \eta ^{\mu \nu}}{z-w}, \nonumber \\
\langle \psi^{\mu}(z) \tilde \psi^{\nu}(\bar w) \rangle & = &
\frac{ ia ^{\mu \nu}}{1-z \bar w},
\label{props}
\end{eqnarray}
where $a^{\mu \nu} = diag \{ -1,1,\ldots,1,-1,\ldots,-1\}$ with
the -1 in the last 9-p entries reflecting the Dirichlet boundary
conditions.
The resulting integrals can be carried out explicitly with
the result
\be
A=f(\varepsilon_i, k_i) 
\frac{\Gamma (1-2s) \Gamma(-t/2)}{\Gamma(1-2s-t/2)},
\label{scatt}
\ee
where the functional form of the 
kinematic prefactor $f$ depends on the type
of string states that scatter off the D-brane \cite{kletho}.
The kinematic invariants are
\begin{eqnarray}
s &=& - {k_1^{\parallel}}^2 =  - {k_2^{\parallel}}^2, \nonumber \\
t &=& 2{k_1^{\parallel}}^2 -2k_1^{\perp} \cdot k_2^{\perp},
\end{eqnarray}
where $ k^\parallel$ and $k^\perp$ are momenta parallel and perpendicular
to the D-brane worldvolume.

The amplitude (\ref{scatt}) is similar to standard string scattering
amplitudes.  This is hardly surprising since the D-brane by
definition interacts with its environment through the open strings
that attach to it.  The poles in the s-channel correspond to an open
string propagating along the D-brane, while the t-channel poles
correspond to a closed string emitted from the D-brane
interacting with the external strings.
The amplitude exhibits the Regge behavior characteristic
of string scattering,
\be
A \sim s^{\frac{t}{2} -1} \approx \frac{1}{s} 
e^{\frac{1}{2} (\log s) t},
\ee
indicating that the transverse thickness of a D-brane, as
seen by strings, is of order the string scale at low
energy and grows with increasing energy of the probe. This
conclusion holds for all Dirichlet p-branes with $p \geq 0$.
For scattering off a D-instanton the kinematics
is degenerate and the form factor exhibits point like
behavior.  

It is not surprising to find that stringy probes cannot see
structure on scales smaller than the fundamental string
length, but since D-branes interact by string exchange
we can also draw some conclusions about high energy
collisions of D-branes.  If the intermediate string
states carry high energy the Regge behavior will take
over, so in order to probe short distances by scattering D-branes
of each other the collisions should be soft and involve
relatively slow moving D-branes.  It turns out that sub-stringy
distances can indeed be probed by slow
D-branes, but we need to introduce a few more tools before
we address that question.

One can also compute mixed closed and open string 
scattering amplitudes in a D-brane background using
more or less standard techniques.
The open string states correspond to excitations
of the D-brane itself and the closed strings
represent a coupling to the rest of the world.
Such amplitudes play an important role in the application of 
D-branes to black hole physics, for instance in
the calculation of Hawking emission from a
near-extremal black hole modeled as a
multi D-brane composite object.  Such
calculations are discussed by Maldacena elsewhere in this 
volume.

\section{D-brane Dynamics}
So far we have only discussed single D-branes at rest.
It is clearly important to develop a dynamical theory
for multiple D-brane configurations
and D-brane motion.  In full generality such a
dynamical description remains a technical challenge.
The formalism we have at our disposal describes
D-branes in terms of a fixed background conformal field
theory and does not generalize in any easy way
to time-dependent backgrounds. In order
to proceed further we have to make approximations. 
Following Bachas \cite{bachas}, we consider the forward
scattering of two parallel p-branes in the eikonal approximation,
where one p-brane moves in a straight line past the other
with impact parameter $b$.
The p-branes interact by string exchange but no back-reaction on
their trajectories or internal states is taken into account.
The calculation involves an amplitude which is a straightforward
generalization of the one-loop open string diagram
we used previously to identify the D-brane tension and \RR\ charge,
with the only difference being that the separation
between the p-branes is no longer constant,
\be
Y^2 = b^2 + v^2 t^2.
\ee
The one-loop amplitude can still be evaluated explicitly
and gives the phase shift for the forward scattering.
The resulting expression is somewhat complicated
and will not be repeated here.  There are two interesting
limits.  One is large $s$, where $s$ is the center-of-mass
energy squared of the scattering process.  
The imaginary part of the phase shift is then \cite{bachas}
\be
Im(\delta) \sim \exp [- \frac{b^2}{\log(s/M^2)}],
\ee
where M is the mass of the  D-brane.
The phase shift exhibits a characteristic length scale,
$b \sim [log(s/M^2)]^{1/2}$,
which grows with energy consistent with the 
Regge behavior of the string scattering
amplitudes considered above.

The other limit is low velocity, $v \ll 1$, where
one finds \cite{bachas}
\be
Im(\delta) \sim \exp [- \frac{b^2}{v}].
\ee
In this case the characteristic length scale
is $b \sim v^{1/2}$ in string units, which is
very small if $v \ll 1$.  This is the first
indication that slow-moving D-branes
can probe distances below the string length.

The two limits involve different values of
the real valued modulus of the annulus
amplitude.  The large $s$ behavior is dominated by
the region in moduli space where the worldsheet
is a long cylinder, i.e. the
closed string channel, whereas the low velocity
amplitude is dominated by the open string
channel where the annulus is long and thin. In
the $v \ll 1$ case, the light degrees of freedom are 
associated with short open strings stretched between
the D-branes.  This suggests an effective
field theory description in terms of
massless open string states, {\it i.e.} a gauge theory,
which turns out to be very useful, as we shall see later on.

The one-loop string amplitude that we have
found so useful can be generalized further
\cite{lif,dkps}. For instance, the two
D-branes do not have to have the same dimensionality.
Consider the forward scattering of a q-brane past a p-brane
with $p \geq q$.  We can only have p-q be an even number
since we are either working in IIA theory where p and q
are both even or IIB theory where they are both odd.
The new feature here is open strings that stretch
between D-branes of different dimensionality,
with some fields that have ND boundary conditions, 
{\it i.e.} Neumann at the p-brane but Dirichlet
at the q-brane.  Once again the amplitude can be 
obtained explicitly but is somewhat messy. It
is useful to introduce
an effective potential as follows
\be
A(v,b^2) = \frac {1}{\sqrt{2 \pi^2}} 
\int_{-\infty}^{\infty} dX^0 V(v,r),
\ee
where $r^2=b^2+(X^0)^2 v^2$. This potential governs
the motion of a non-relativistic q-brane interacting with a
p-brane with p$\geq$q.  The low velocity behavior
depends on the value of p-q. 

For p=q we recover the case we discussed above and the
potential takes the form
\be
p=q: \>\> V(v,r) = - \frac {a v^4}{r^{7-p}}
 \Bigl(1 + O \bigl( \frac {v^2}{r^4} \bigr) \Bigr),
\ee
where a is a positive constant.
The velocity dependence means that identical D-branes
at rest feel no force, consistent with the BPS property.
At non-zero velocity the system no longer has unbroken
supersymmetry and there is a net attractive force between
moving D-branes.  The expansion parameter at
low velocities is $v^2/r^4$ which again suggests
a characteristic length scale $r_0 \sim v^{1/2}$.

For $p=q+4$ there is still no force between the
D-branes when they are at rest, but the
velocity dependent potential is
different from the p=q case,
\be
p=q+4:\>\>  V(r,v) = - \frac {\tilde a v^2}{r^{3-q}}
 \Bigl(1 + O \bigl(\frac {v^2}{r^4}\bigr) \Bigr),
\ee
The leading term goes as $v^2$, so it is natural to
absorb it into the kinetic term
\be
K \sim \frac {1}{g} \bigl( 1 - \frac {\tilde a g}{ r^{3-q}} 
\bigr) v^2 .
\label{kin}
\ee
The low velocity scattering can then be described in terms
of geodesic motion on a moduli space with a
non-trivial metric \cite{dkps}.

Another case of interest is $q=\bar p$, the scattering
of a Dirichlet p-brane and anti-p-brane.
These objects do not satisfy a BBS condition even at
rest and one expects a nontrivial potential already at v=0.
There is in fact an instability if the p-brane and antibrane
get too close to each other \cite{bansus}.  To see this,
consider the one-loop amplitude (\ref{vacamp}) for identical
p-branes, with a small but critical difference.
In the present case the objects have opposite sign \RR\
charge and the corresponding term, $F_3$ in the square
brackets in (\ref{vacamp}) enters with the
opposite sign. Now there is no longer a 
cancellation but instead the terms add and
the leading behavior of the integrand for large
t is
\begin{eqnarray}
A &=& V_{p+1} \int_{0}^{\infty} \frac{dt}{t} 
(8 \pi^2 t) ^{- \frac {p+1}{2}} \nonumber \\
& & \times e^{(\pi - \frac {Y^2}{2 \pi}) t } 
\bigl(1 + O (e^{-2 \pi t}) \bigr).
\end{eqnarray}
The integral diverges for $Y^2 < 2 \pi^2$. The system is unstable
against $p - \bar p$ annihilation at short distance.  The actual
annihilation process is presumably complicated
and results in the emission of a large number of
closed strings.  

There is a similar instability in the amplitude with
$p = q+2$.  In this case the system contains two
D-branes of different dimensionality which do not
annihilate each other
but form a bound state with non-vanishing
binding energy.

\section{Effective Action for D-Branes}
We can generate an effective action to
describe low-energy
processes involving D-branes by the usual string theory
procedure.  The first step is to identify the
appropriate two-dimensional worldsheet
theory with general couplings. Since
D-branes are associated with open
strings it is the possible boundary
couplings that are of interest here,
\begin{eqnarray}
S_b & =& \int ds \bigl[  \sum_{m=0}^{p} 
A_m(X^0, \ldots, X^p) \partial_t X^m \nonumber  \\
& & + \sum_{i=p+1}^{q} \nu_i (X^0, 
\ldots, X^p) \partial_n X^i \bigr],
\label{bact}
\end{eqnarray}
where $\partial_t$ and $\partial_n$ denote the tangential and normal
derivatives at the worldsheet boundary.  
The presence of the boundary
action (\ref{bact}) leads to a generalization of the mixed
Neumann/Dirichlet boundary conditions for the
open strings attached to the D-brane.  The above
terms are the most general boundary action of scaling dimension
one ({\it i.e.} renormalizable) for scalar fields 
$X^{\mu}, \mu = 0, \ldots, 9$.  
The function $A_m$ is interpreted as a $U(1)$
gauge potential that lives on the worldvolume and $\nu_i$
represents a shift in the D-brane position in the transverse
directions.
We have not
written the accompanying fermion terms required for
worldsheet supersymmetry. The form of the boundary action
(\ref{bact}) is constrained by
the Dirichlet boundary condition in the free
theory in that $A_m$ and $\nu_i$ are only functions
of the worldvolume embedding coordinates.

The next step towards the effective action is
to couple these boundary terms to the general
non-linear sigma model in the bulk of the
string worldsheet and require the combined
system to have conformal symmetry
in order to decouple negative norm
states from the string spectrum.  This 
places a set of restrictions on the coupling 
functions of the theory which can be cast as
dynamical field equations in the ten-dimensional
target space. The final step is to
identify a target space action functional
whose equations of motion are
equivalent to the field equations that
follow from two-dimensional conformal 
invariance.  This procedure was carried out
for D-branes in bosonic
string theory by Leigh \cite{leigh} and
the resulting D-brane effective action
is a generalization of the Born-Infeld
action of non-linear electrodynamics,
\begin{eqnarray}
S_p &=& - T_p {\int} d^{p+1} \sigma e^{- \phi} \nonumber \\
& &\times \sqrt{\det (g_{\mu \nu} + b_{\mu \nu} + 2 \pi F_{\mu \nu})},
\label{bia}
\end{eqnarray}
where the coefficient $T_p$ is proportional to the
D-brane tension, $g_{\mu \nu}, b_{\mu \nu}$ and $\phi$ are spacetime
fields pulled back to the p-brane worldvolume, and
the components of the gauge field strength are
\begin{eqnarray}
F_{mn}& = & \partial_m A_n - \partial_n A_m, 
\>\>\> m,n=0,\ldots,p \nonumber \\
F_{ij} & = & 0, 
\>\>\> i,j=p+1, \ldots,9, \nonumber \\
F_{mi} & = & \partial_m \nu_i .
\end{eqnarray}
The $F_{mn}$ components are the worldvolume gauge
field strength while the $F_{mi}$ components
reflect the D-brane velocity and tilt in the transverse
directions.  The full effective action also includes the
usual ten-dimensional action for $g_{\mu\nu}$, $b_{\mu\nu}$, and
$\phi$.

In type II superstring theory the \NS\ sector 
contributes a Born-Infeld term identical
to (\ref{bia}) and in addition
the \RR\ sector gives rise to Chern-Simons type
terms \cite{lido,many} that can be neatly summarized as
\be
S'_p = i \mu_p \int_{V_{p+1}} a \wedge Str \,e^{(b+2 \pi F)},
 \label{rra}
\ee
where $a = \sum_n a^{(n)}$ is the sum of all the \RR\ gauge
potentials in type II theory.  The exponential
in (\ref{rra}) also involves a sum of terms but the
integral picks out forms of overall rank p+1 from the 
product.
The couplings between the worldvolume gauge field
and the \RR\ sector potentials play an
important role in the study of D-brane bound
states \cite{pol2,lido}.  Consider for example a p-brane
with a non-vanishing gauge field strength $F_{\mu\nu}$ in its
worldvolume.  According to (\ref{rra}) such a configuration carries 
(p-2)-brane charge in addition to its p-brane charge.  
It turns out to have lower energy than the p- and (p-2)-branes
separately and is in fact a realization of the (p,p-2) bound state
mentioned at the end of the previous section.

We have only kept track of the bosonic
part of the D-brane action but there are of
course also fermionic terms as required
by supersymmetry.
The supersymmetric generalization of the 
abelian Born-Infeld action (\ref{bia}) has
been written down \cite{cederwall}.
It is rather involved and will not be needed
here. 

The $U(1)$ worldvolume
gauge field is associated with massless
open strings which end on the D-brane.
An interesting new feature appears when we consider
multiple D-brane configurations.            
If some of the D-branes coincide then there are additional 
massless open strings that go between different D-branes. When
the D-branes are well separated such strings have a large
mass due to their tension and are suppressed at
low energies. When $N$ D-branes are on top of each
other there are $N^2$ massless open string modes
that fill out the adjoint representation of $U(N)$ and
the worldvolume gauge symmetry is
enhanced from $U(1)$ to $U(N)$ \cite{pol3,witten}.

The D-brane position in the transverse directions, $\nu_i$, is
also promoted to a $U(N)$ matrix.  This has played
a crucial role in exploring D-brane dynamics.
It indicates that spacetime coordinates are 
replaced by  non-commuting variables when we get
down to very short distances. 

This formalism also leads to a nice geometric picture
for symmetry breaking when a gauge theory
can be realized as the worldvolume theory
of a collection of D-branes.  The unbroken phase
corresponds to all the D-branes coinciding
and then the gauge symmetry can be partially
or completely broken by moving some
or all of the D-branes apart. 

The non-abelian generalization of the
D-brane effective action remains a subject of
investigation at the present time.  In particular
the generalization of the
non-linear Born-Infeld action is
troublesome \cite{tseytlin}.
In flat spacetime when the gauge field
strength is weak and the D-branes are all
moving slowly the
non-linear terms in the D-brane action
will be small and the effective action is well
approximated by the appropriate dimensional
reduction of D=10,
supersymmetric, $U(N)$ Yang-Mills
theory \cite{witten}.
Most studies of D-brane dynamics to date have
been based on the Yang-Mills approximation
rather than the full non-linear theory and we
will describe an example in the following section.

The dimensionally reduced Yang-Mills theory
will in particular include a potential term
for the matrix valued D-brane positions which is
proportional to
$Tr([\nu_i, \nu_j][\nu^i, \nu^j])$
and comes from the $F_{ij}$ components of the non-abelian
field strength.  The D-branes can only be separated along flat directions
of this potential but these occur precisely
when the $\nu^i$ all commute and the gauge
group is broken from $U(N)$ to $U(1)^N$, as
expected from considering open strings attached to the D-branes.

\section{D-Particle Scattering}
As a simple application of the above ideas we
briefly consider the scattering of two non-relativistic
0-branes \cite{dkps,dfs,kabpou}. This involves
the dimensional reduction of D=10 supersymmetric $U(2)$
gauge theory to a 0+1 dimensional theory on the 
D-particle worldline.  The $U(2)$ symmetry factorizes
into $U(1) \times SU(2)$ in a natural way,
where the $U(1)$ factor represents the center-of-mass
degrees of freedom and the interesting dynamics is in the 
$SU(2)$ sector.

The worldline action including the fermionic
terms is
\be
S =  \int dt \bigl[ \frac {1}{g_s} Tr (F_{\mu \nu} F^{\mu \nu}) 
 - i Tr (\bar \psi \Gamma^\mu D_{\mu} \psi)\bigr],
\ee
where $A_0$, $\nu_i$, and $\psi$ take values in the adjoint
representation of $U(2)$ and
\begin{eqnarray}
F_{0i} & =  & \partial_0 \nu_i + [A_0, \nu_i], \nonumber \\
F_{ij} & = & [\nu_i, \nu_j], \nonumber \\
D_0\psi & = & \partial_0\psi + [A_0, \psi], \nonumber \\
D_i \psi & = & [\nu_i, \psi].
\end{eqnarray}
A simple scaling argument reveals the characteristic length scale
of the dynamics \cite{kabpou}.
In terms of the rescaled variables $\tilde t = g^{1/3} t$, 
$\tilde A_0 = g^{-1/3} A_0$,
and $\tilde \nu_i = g^{-1/3} \nu_i$, the action takes
the form
\be
S= \int d\tilde t \bigl[Tr (\tilde F_{\mu \nu} \tilde F^{\mu \nu}) 
 - i Tr (\bar \psi \Gamma^{\mu} \tilde D_{\mu} \psi)\bigr],
\ee
where the string coupling has been scaled out.  The
characteristic length scale is therefore $l \sim g^{1/3} l_s$,
which is in fact the Planck length of the
eleven-dimensional supergravity theory which is
the strong coupling dual of type IIA string theory.
At any rate, this length scale is short compared to
the fundamental string length $l_s$ when the string
coupling $g_s$ is weak, and this is an
indication
that slow moving D-branes can probe distances below
the string length.  We will not
present a detailed study of the D-particle system
here, but we note that the dimensionally
reduced Yang-Mills theory gives rise to
a velocity dependent potential $V \approx a v^4/r^7$ in agreement
with our earlier considerations based on the one-loop open string
amplitude.  String duality requires there
to be a bound state at threshold in this system.
The existence of such a bound state has recently
been proven \cite{setste}. This required some
fairly heavy
mathematical machinery but relatively simple
arguments can be given \cite{dkps,dfs} that
in such a bound state the D-particles
would be separated by $r \sim g^{1/3}_s$ and have velocities
$v \sim g^{2/3}_s$, which is consistent with the
relation $r \sim v^{1/2}$ obtained from the one-loop string
amplitude.

\section{T-Duality and D-Branes}
Our final topic is T-duality in type II string 
theory and the transformation 
properties of D-brane configurations
under this symmetry.  D-branes were in fact
originally discovered when investigating how
T-duality acts on open strings \cite{dlp}.

Consider the closed string mode expansion
$X^{\mu}(z, \bar z) = x^{\mu}(z) + \tilde x^{\mu}(\bar z)$, 
where $z= e^{\tau -i\sigma}$ and
\begin{eqnarray}
x^{\mu}(z) =x_0^\mu +i(-\alpha_0^{\mu} log z 
+ \sum_{m \neq 0} \frac {1}{m} \alpha^\mu_m z^{-m}) \nonumber \\
\tilde x^{\mu}(\bar z) = \tilde x_0^\mu 
+i(-\tilde \alpha_0^\mu log \bar z 
+ \sum_{m \neq 0} \frac {1}{m} \tilde \alpha^\mu_m \bar z^{-m}) 
\label{modes}
\end{eqnarray}
The spacetime momentum is the sum of the left- and
right-moving zero modes,
\be
\alpha_0^{\mu} + \tilde \alpha_0^\mu = 2 p^\mu ,
\ee
and their difference gives the change in
$X^{\mu}(z, \bar z)$
upon going around the closed string,
\be
\alpha_0^{\mu} - \tilde \alpha_0^{\mu} = \frac {1}{2 \pi}
\bigl(X^{\mu}(\sigma + 2 \pi, \tau) 
- X^{\mu}(\sigma, \tau)\bigr).
\ee
For non-compact spacetime directions $X^{\mu}$ must be
single valued and it follows that
$\alpha_0^{\mu} = \tilde \alpha_0^{\mu} = p^{\mu}$.
If, on the other hand, one of the coordinates, say $X^9$,
parametrizes a circle of radius R, then $X^9$
only has to come back to itself up to
an integer multiple of $2 \pi R$ when
$\sigma \rightarrow \sigma + 2 \pi$,
\be
\alpha_0^9 - \tilde \alpha_0^9 = mR.
\label{adiff}
\ee
The momentum in the
compact direction takes discrete values $p^9 = n/R$ so
that
\be
\alpha_0^9 + \tilde \alpha_0^9 = \frac{2n}{R}.
\label{asum}
\ee
The left- and right- moving zero modes in the
compact directions are then
given by
\begin{eqnarray}
\alpha_0^9 &=& (\frac{n}{R} + \frac{mR}{2}), \nonumber \\
\tilde \alpha_0^9 &=& (\frac{n}{R} - \frac{mR}{2}).
\end{eqnarray}

The mass spectrum and interactions of the string theory
is invariant under the T-duality transformation
\be
T_9:  R \rightarrow \frac{2}{R},  m \longleftrightarrow n.
\ee
In other words, the theories at radius
R and $2/R$ are exactly the same
when the winding and momentum quantum
numbers are interchanged.  The T-duality transformation
can be written
\be
T_9: X^9(z, \bar z) \rightarrow X'^9(z, \bar z) = x^9(z)-\tilde x^9(\bar z),
\ee
{\it i.e.} as a parity transformation acting on right movers
only.  The corresponding transformation on the 
spin fields is
\be
T_9: S^\alpha \longleftrightarrow S^\alpha, 
\quad \tilde S^\alpha  \longleftrightarrow
\tilde S^{\dot{\alpha}}.
\ee
The chirality of the right-moving spin field is changed
under $T_9$ which means that the duality
transformation relates IIA and IIB theory.
This remains true when T-duality acts on 
an odd number of compact dimensions
while acting with T-duality on
an even number of dimensions
does not change the chiral character of
the theory.

\vskip 0.5cm
\vbox{
{\centerline{\epsfxsize=1.0in \epsfbox{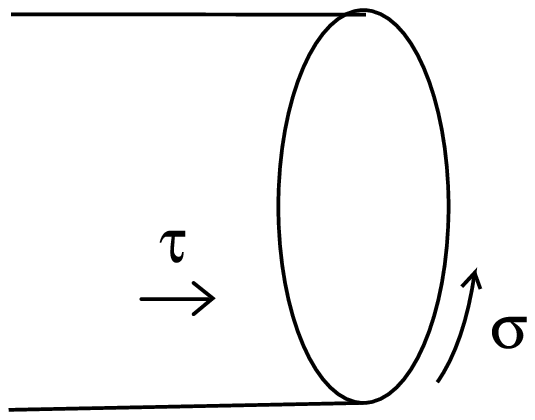}}}
{\centerline{ FIGURE 4:
Worldsheet boundary in }}
{\centerline{closed-string channel.}}
}
\vskip .5cm
In the open string sector T-duality acts on
the boundary conditions in non-trivial
way.  To see this it is convenient to
view the worldsheet boundary in a
closed string channel, as
shown in Figure~4.  
This can always be
arranged by a conformal transformation
and choice of parametrization.  The
Neumann boundary condition,
$\partial_\tau X^\mu = 0$, written in
terms of complex coordinates
$z = e^{\tau - i\sigma}$, is
\begin{eqnarray}
0 &=& (z \partial_z + \bar z \partial_{\bar z}) X^\mu (z, \bar z)
\nonumber \\
&=& z \partial_z x^\mu (z) + \bar z \partial_{\bar z} \tilde x (\bar z).
\end{eqnarray}
Similarly, the Dirichlet condition, $\partial_\sigma X^\mu = 0$,
becomes
\be
0=z \partial_z x^\mu(z) 
- \bar z \partial_{\bar z} \tilde x^\mu (\bar z).
\ee
When T-duality acts on a compact coordinate
the right-moving part of that coordinate changes sign and
thus Neumann and Drichlet boundary conditions are
exchanged.  Now act with a duality transformation $T_\mu$ on
a Dirichlet p-brane with worldvolume along $(X^0, \ldots, X^p)$.
If $\mu \leq p$ the dimension
of the D-brane is reduced by one, 
while it is increased by one if $\mu > p$.  This
is consistent with the observation that T
duality on an odd number of coordinates
changes IIA theory into IIB and {\it vice versa}.

T-duality can often be used to relate problems that might at
first sight seem quite different.  For example, bound
states of Dirichlet p- and (p-n)-branes for a given value of
n with different values of p can all be mapped into each other
by T-duality transformations. When T-duality is combined with the
strong/weak coupling $SL(2,Z)$ duality of IIB theory the 
class of related problems is further enlarged.

\section{Parting Words}

D-branes are interesting and useful objects.  They have enhanced
our understanding of non-perturbative string theory and continue
to find new applications.  There is already a vast literature on
D-brane physics and it seems likely they will continue to
play an important role in the future development of string theory.
In these lectures we described how D-branes are introduced into
type II superstring theory and discussed some of their dynamical
properties.  We only scratched the surface of the subject and many
important topics were left out.  For example, we did not cover
unoriented string theory at all, where a lot of interesting work
involving D-branes has been done.  We nevertheless hope that the
reader comes away with some impression of the power and beauty
of D-brane technology.

The author thanks the organizers of the 33rd Karpacz Winter School 
for the opportunity to lecture and for their kind hospitality.  
He also thanks A. Peet for helpful comments on the manuscript.
This work was supported in part by a US Department of Energy 
Outstanding Junior Investigator Award, DE-FG02-91ER40671.

\end{document}